\documentclass[10pt,journal,compsoc]{IEEEtran}
\usepackage{amsmath,amssymb,amsfonts}
\usepackage{algorithmic}
\usepackage{graphicx}
\usepackage{textcomp}
\usepackage{braket}
\usepackage[ruled,vlined]{algorithm2e}
\usepackage{graphicx}
\usepackage{makecell}
\usepackage{caption}
\usepackage{amsthm}
\usepackage{multirow}
\usepackage{hhline}

\ifCLASSOPTIONcompsoc
  \usepackage[nocompress]{cite}
\else
  \usepackage{cite}
\fi
\ifCLASSINFOpdf
\else
\fi

\begin{document}
%
\title{QDLC - The Quantum Development Life Cycle}
%
%
%
%

\author{Nivedita~Dey,~\IEEEmembership{QRDLab India,~University of Calcutta,}
		Mrityunjay~Ghosh,~\IEEEmembership{QRDLab India,~University of Calcutta,}
        Subhra~Samir~kundu,~\IEEEmembership{QRDLab India}
        and~Amlan~Chakrabarti,~\IEEEmembership{University of Calcutta}
\IEEEcompsocitemizethanks{\IEEEcompsocthanksitem N. Dey is Senior researcher and Head of consulting in Quantum Computing domain of QRDLab, Kolkata and is with University of Calcutta.\protect\\
E-mail: nivedita@qrdlab.com
\IEEEcompsocthanksitem M Ghosh is Head of Research in QRDLab, Kolkata and is quantum computing researcher at University of Calcutta.
\IEEEcompsocthanksitem S S Kundu is with QRDLab and Amity University Kolkata.
\IEEEcompsocthanksitem A Chakrabarti is  with A. K. Choudhury School of Information Technology, University of Calcutta.}
\thanks{Manuscript submitted 16 October, 2020.}}

%
%

\markboth{QDLC - The Quantum Development Life Cycle,~Vol.~43, No.~18, October~2020}%
{Dey \MakeLowercase{\textit{et al.}}: QDLC - The Quantum Development Life Cycle}
%



\IEEEtitleabstractindextext{%
\begin{abstract}
The magnificence grandeur of quantum computing lies in the inherent nature of quantum particles to exhibit true parallelism, which can be realized by indubitably fascinating theories of quantum physics. The possibilities opened by quantum computation (QC) is no where analogous to any classical simulation as quantum computers can efficiently simulate the complex dynamics of strongly correlated inter-facial systems. But, unfolding mysteries and leading to revolutionary breakthroughs in quantum computing are often challenged by lack of research and development potential in developing qubits with longer coherence interval, scaling qubit count, incorporating quantum error correction to name a few. Putting the first footstep into explorative quantum research by researchers and developers is also inherently ambiguous – due to lack of definitive steps in building up a quantum enabled customized computing stack. Difference in behavioral pattern of underlying system, early-stage noisy device, implementation barriers and performance metric cause hindrance in full adoption of existing classical SDLC suites for quantum product development. This in turn, necessitates to devise  systematic  and cost-effective techniques to quantum software development through a Quantum Development Life Cycle (QDLC) model, specifying the distinguished features and functionalities of quantum feasibility study, quantum requirement specification, quantum system design, quantum software coding and implementation, quantum testing and quantum software quality management.
\end{abstract}

}

\maketitle

\IEEEdisplaynontitleabstractindextext

%
\IEEEpeerreviewmaketitle

\section{Introduction}
In today’s machine age even with the advent of Artificial Intelligence (AI) and
Machine Learning (ML), machines cannot be brought any where close to kind
of real intelligence, that a human brain possesses. The grammar or syntactic
structure of a machine adopts deterministic approach, which works hand in hand
with a pre-defined algorithm, simple or a complex one, to solve a particular task.
But the way a human mind takes data coming through our sensors and finds
structure about the outer world follows pretty sophisticated statistical
inferencing, backed up by probabilistic reasoning. Thus, it depicts an obvious fact
of a human brain being more powerful in comparison to that of an artificial system.
The difference does not lie in the processing power of the two systems, it is in the
way two systems process and interpret a data to reach a conclusion. This feature of
exhibiting chances of exploring multiple choices by a powerful technique is called
non-determinism.
Quantum computing exhibits non-determinism by taking the advantage of laws of
quantum mechanics found in bewilderly random nature and represents a
fundamental change from classical information processing. \cite{basicqc} Today’s classical, transistor-based machines utilize classical bits, 0 and 1 to represent an
information. In quantum setting, particles do normally exist as several
mathematical possibilities rather than one actual object in absence of any
observation. Like classical bit, the basic quantum processing unit is termed as a qubit or quantum bit but can depict a “0”, a “1” or any combination of them, unlike its classical counterpart.

Even if one is able to make very high-quality qubits, creating and making use of these quantum computers (QCs) brings a new set of challenges. They use a different set of operations than those of classical computers, requiring new algorithms, software, control technologies and hardware abstractions.

Methods to debug quantum hardware and software are of critical importance. Current debugging methods for classical computers rely on memory and the reading of intermediate machine states, which is computationally infeasible in a quantum computer. A quantum state cannot simply be copied for later examination and any measurement of a quantum state collapses it to a set of classical bits, bringing computation to a halt. \cite{measurement} Building a useful device is much more complex than just creating the hardware tools those are needed to create and debug QC-specific software. Since quantum programs are different from programs for classical computers, new approaches to debugging are essential for the development of large-scale quantum computers. Research and development is needed to further develop the software tool stack. Since, contemporaneous development of the hardware and software tool chain will shorten the development time, intrinsic complexity and development cost, overall progress is required to develop a scalable quantum hardware with a suited model of computation. \cite{qcomplexity}

This in turn necessitates the introduction of a new model of Quantum Development Life Cycle (QDLC) unlike classical Software Development Life Cycle (SDLC), where a pre-defined pathway for implementing and solving large projects on quantum both in a time-efficient and resource-efficient manner. The rest of the paper starts with the literature review section in section 2. Section 3 covers the basics of motivation behind the development of classical SDLC models and a brief introduction of classical waterfall model and its variations. Section 4 illustrates the introduction to Quantum Development Life Cycle as Quantum Waterfall model. Section 5 and 6 shed light on quantum feasibility study and requirement analysis of quantum hardware as well as software. Section 7 elucidates several design aspects of quantum system followed by a quantum coding and implementation stage specified in section 8. Section 9 describes quantum testing through quantum state reconstruction along with the quantum verifier and debugging tools. In section 10, Quantum Software Quality Management (QSQM) has been discussed where optimization techniques are specified which can drastically affect the performance of quantum algorithm by reducing the estimated runtime and other computational overhead. In section 11, the near-term quantum solutions and quantum computing market barriers have been illustrated. The last section specifies the conclusion and scope for future research growth on related areas. Here our proposed model is not to reconstruct the entire history of quantum hardware and software development but to provide an agile path to quantum computational development and further studies.

\section{Literature Survey}
Leveraging the true potential of quantum hardware and software needs development of conceptual framework like QDLC model for development of quantum software. In this section we have presented a comprehensive review of existing literature in both classical SDLC models and quantum software engineering, which are illustrated throughout the paper in respective contexts. Herbert D. Benington had first proposed the classical waterfall model as a SDLC model and identified its stages in 1956, \cite{benington} where each stage of the waterfall model can be visited only once and there is no provision to revisit a preceding state from a successive state. This limitation of classical waterfall model was overcome in Royce's paper in 1970 where the concept of a feedback loop was introduced to reduce rigidity in classical waterfall. \cite{royce} Classical SDLCs are suited for classical computing relying on extended Church-Turing theory. In order to make faster execution of a computational problem in classical context, the aim is to reduce the time required  and steps involved in implementation. True potential of quantum computer was harnessed when Peter Shor first had discovered a quantum algorithm for prime factorization to achieve exponential speedup over classical algorithm. \cite{shor}  Another quantum computing milestone was the discovery of a quantum unstructured database search algorithm by L. K. Grover \cite{grover}, where he had shown how quantum search algorithm can outperform classical unstructured search algorithm and achieve quadratic speedup. The primary computing paradigms of quantum algorithms can be classified as Quantum Fourier Transform (QFT), Quantum Amplitude Amplification (QAA), Quantum adiabatic algorithms, Quantum simulation algorithms and Quantum walk algorithms, \cite{qalgo} \cite{qaa} operating over two different models of quantum computing, namely gate model and adiabatic model. These quantum algorithms can be logically realized with the help of quantum circuits which incorporate various single-qubit and multi-qubit quantum gates and their implementation involves decomposition of complex quantum gates into a cascade of simple one-qubit and two-qubit quantum gates using Fault Tolerant Quantum Logic Synthesis (FTQLS). \cite{ftqls} But, since quantum bits suffer from gradual environmental decoherence due to amplitude and phase damping, it is necessary to make two interacting qubits adjacent to each other. Lin et. al. referred it as Physical design Aware Quantum Circuit Synthesis (PAQCS). \cite{paqcs} Niemann et. al. \cite{pmd} had proposed the approach to synthesize different quantum circuits for dedicated Physical Machine Descriptions (PMDs) based on underlying hardware.

The design aspect of quantum programming is to build an executable program which can be run on a quantum computer. Quantum operations are performed on quantum register of qubits to encode the problem state space and classical register to obtain the final measurement, once a quantum state is collapsed into classical state. \cite{measurement} The model of Quantum Turing Machine (QTM) was first proposed by Deustch. \cite{qtm} Later, a model of a Quantum Random Access Machine (QRAM) for programming language was proposed by Knill, where a classical-controlled quantum system was introduced. \cite{qram} The first practical quantum programming language QCL was introduced in 1988 by \"Omer, which followed a C-like syntax. \cite{qcl} Later on many other quantum programming languages and tools had been designed like QISKIT \cite{qiskit}, Cirq \cite{cirq}, Q\# \cite{qsharp}, PyQuil \cite{pyquil} to name a few significant ones. Another quantum language which is used to implement experiments with low depth quantum circuits, is Open Quantum Assembly (OpenQASM) language. \cite{qasm} This language is a framework for intermediate representation of quantum instructions and has similar qualities to traditional hardware description languages like Verilog.

Despite the technological advancements, progress on R\&D of quantum hardware and quantum software is still very much in its nascency. Today's available quantum devices operate on Noisy Intermediate Scale Quantum (NISQ) computing model. \cite{nisq} The applications those are likely to require fewer number of qubits are only suitable to be performed in NISQ devices. \cite{nisq} The commercial availability of a Fault Tolerant Quantum Computing (FTQC) model is years to achieve \cite{ftqc}, creating potential challenges in near-term quantum era. \cite{challengeqc} Qubit performance often gets adversely affected due to quantum error and hence causes hindrance in achieving scalability of quantum devices. \cite{qecc} \cite{scalability} \cite{qubitquality} Trade off in computing is measured by optimizing the underlying architecture in terms of improved compiler optimizations, efficient quantum algorithms and quantum errors in terms of several coupling issues. \cite{qecc} \cite{coherencequbit} 
 
\section{Potential drivers behind ideation of classical SDLC models}
\begin{figure}
  \includegraphics[width=\linewidth]{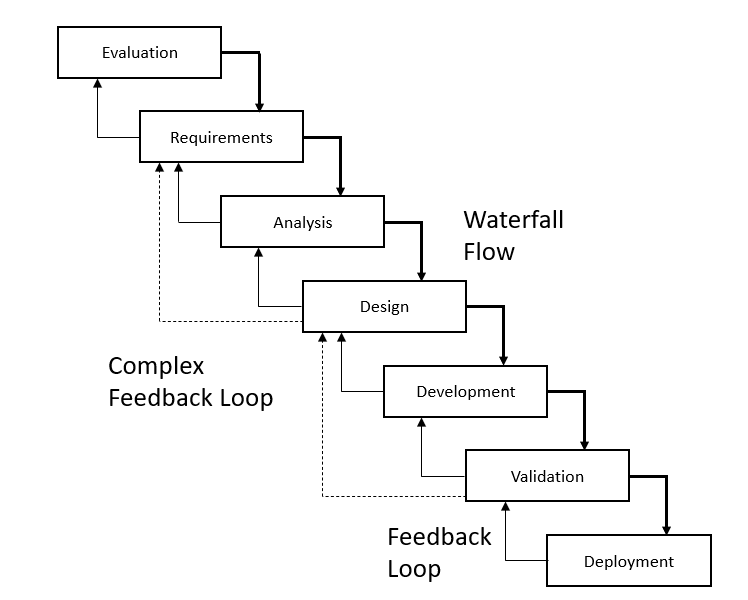}
  \caption {Royce’s Model for Software Development}
  \label{fig:roysmodel}
\end{figure}

SDLC is a conceptual framework or process which characterizes the structure of the stages involved in the development of an application from the very beginning of its initial feasibility study through to its deployment. It is a graphical model depicting the different phases through which a software evolves and usually accompanied by a textual description of different activities that needs to be carried out in each phase.
\begin{figure*}[t]
  \includegraphics[width=\linewidth]{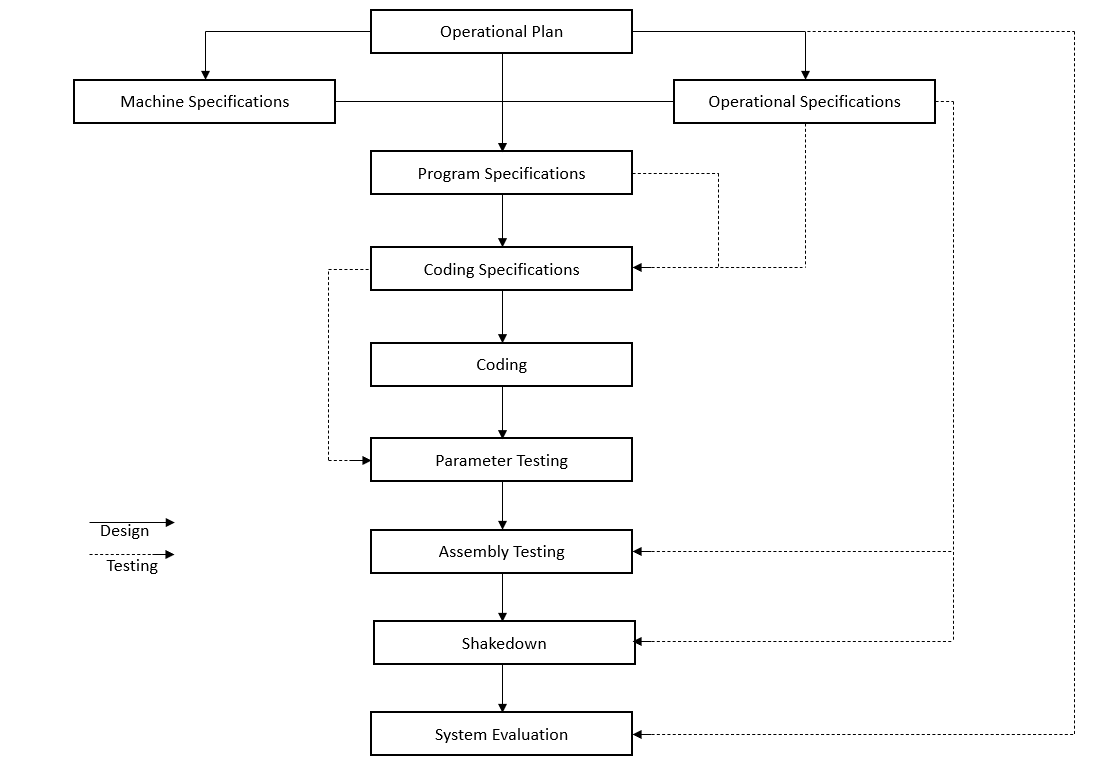}
  \caption {Bennington’s Model for Software Development}
  \label{fig:Bennington}
\end{figure*}
Conceptualization and formulation of a QDLC model necessitate the researchers to understand the root cause and framework behind development of classical SDLC models.
In the first two decades of computer era around 1950s and 1960s, the primary focus of computation was on mainframe computers. Minicomputers came into scene in mid 1960s, moreover IBM PC and Macintosh had no footsteps before early to mid-1980s. Software development process was at a naive stage then, with issues of resource scarcity like computing speed, available memory and system design impacting software operability. The development environment was in nascency with availability of low-level programming languages like assembly, ALGOL, Fortran, COMTRAN, FLOW-MATIC, COBOL and PL/I, which in turn was susceptible to coding error due to lack of conciseness. Modern Software development tools like code-highlighting, robust Integrated Development Environment (IDE), version control systems like Git and high-end function-oriented and object-oriented programming languages did not exist in early stage of classical computation era. Moreover, there was lack of domain specific expertise as early software developers were mostly engineer or mathematicians. Hence, the style of programming adopted in early stage was more of a ‘code and fix’ nature. The concepts of structured programming with use of sub-routines, block structures, loops were not feasible before late 1960s. The most notable difference between modern commercially available computers and early computers was that of cost centers as cost of developers, project staffs, computing hardware and other resources were far incomparable to that of current scenarios. Just to have a glimpse of then available architecture and its cost, a GD ERA 1103 computer costs around \$600 per hour. This implied to an increase in effort to perform analysis, coding and quality testing in offline mode as computing time was too expensive to be used in an unplanned manner. The last contextual factor to be considered for the development of SDLC models was project scaling as expanding or replicating pilot or small-scale projects to broaden the effectiveness of an intervention requires diffusion, dissemination and implementing innovation overcoming the massive complexities like planning, requirement elicitation, use of appropriate metrics, risk management to name a few. The thrive in cost control, quality management and project management had led to development of the first stagewise architecture of classical waterfall model in 1956 by Herbert D. Bennington as described in figure \ref{fig:Bennington}. Later, Winston W. Royce in 1970 recognized the unforeseen design difficulties when a base line is created at the end of each stage and hence enhanced the basic waterfall model by providing a feedback loop so that each preceding stage could be revisited. But this arrangement might be proven insufficient when an iteration needs to be transcended the iteration path of the succeeding-preceding stages. Hence, Royce proposed a complex feedback loop to minimize the risk of revisiting a stage several times as shown in figure \ref{fig:roysmodel}.

Although, there is an abundance of classical SDLC models in existence, we have mentioned only the very basic classical and iterative waterfall models as the prerequisite of our proposed QDLC model as these two models had underpinned all other models by creating a firm foundation for requirements to be defined and analyzed prior to any design and development.
\section{Quantum Development Life Cycle (QDLC)}
\begin{figure*}[t]
  \includegraphics[width=\linewidth,height=12cm]{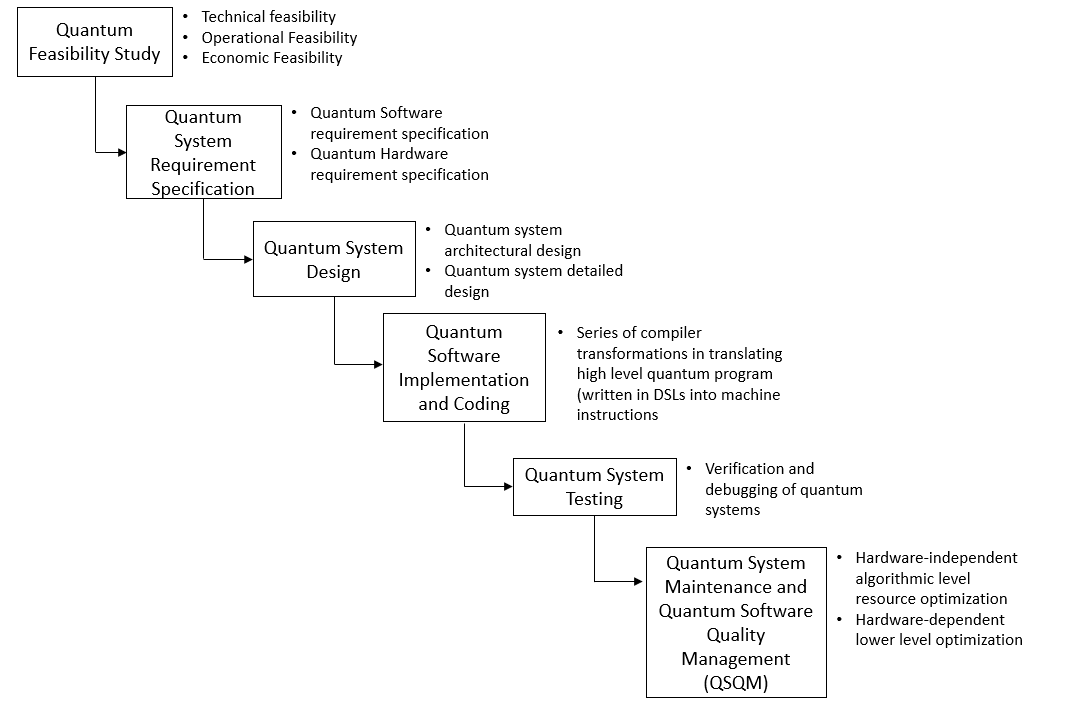}
  \caption {Stage-wise Quantum Development Life Cycle (QDLC) model ('Quantum Waterfall Model')}
  \label{fig:qdlsstage}
\end{figure*}

A full-scale technology drift to bring a paradigm shift from classical computing to quantum computing requires to adopt a set of well-defined engineering tasks. Like early stages of classical computation era, quantum computing is in its infancy as commercial availability of a full-scale quantum hardware with sufficient qubit support in terms of quality and count is unforeseen in the future. Insufficiency of adequate quantum hardware and software resources, inherently erroneous quantum-mechanical system and cost-intensive quantum hardware are the primary concerns in the development of a quantum system. In order to suffice the thrive for a systematic, cost-effective techniques meeting a relatively high level of certainty, a quantum development life cycle model (QDLC) is proposed. The model of QDLC is inspired by classical waterfall model, where a stage-wise architecture is shown in figure \ref{fig:qdlsstage} to depict the sequence of all the stages required in the quantum development process starting from conceptualization and ideation to implementation.
In our proposed approach, we have identified the ‘Quantum Waterfall’ model to undergo the following stages.
\begin{itemize}
\item Quantum Feasibility Study
\item Quantum Requirement Specification
\item Quantum System Design
\item Quantum Software Implementation and Coding
\item Quantum System Testing
\item Quantum System Maintenance and Quantum Software Quality Management (QSQM)
\end{itemize}

\begin{itemize}
\item \emph{Quantum Feasibility Study :} While a classical SDLC model performs technical feasibility and financial feasibility under feasibility study, quantum feasibility study will additionally incorporate analyzing the operational feasibility of the product development. The quantum feasibility study activity involves analysis of the problem statement and collection of all relevant information regarding the product such as input data to the system, processing required to be carried out on those data, output data to be produced by the system, as well as several behavioral constraints relating to the system. In quantum feasibility study, the first step involved is data collection which is followed by data analysis to derive an abstract problem definition, strategy formulation and evaluation in order to make a high-level decision.
\item \emph{Quantum Requirement Specification :} In quantum system requirement specification, requirement analysis activity is carried out to weed out all the incompleteness and inconsistencies in the requirements gathered. In quantum setting, quantum software requirements specification consists of different classical and quantum software supports like quantum tools, logical circuit synthesizers, classical validator modules. Quantum hardware requirements specification is used to specify the underlying hardware requirements like number of qubits required, available quantum volume, classical hardware processors and many more. Additionally, quantum error threshold associated with underlying processing logic must be explicitly specified in accordance to the hardware  and software requirement specification. \cite{qubitquality}
\item \emph{Quantum System Design :} In classical context, the goal of a design phase is to transform the requirements specified in requirement specification document into a structure which is suitable for implementation in some classical programming languages. Quantum system design involves deriving the software architecture through quantum system architectural design and detailed design. Architectural design focuses on functionalities and inter-relationships among quantum hardware and software components, whereas detailed design deals with quantum algorithms, processing logic and data definition semantics. 
\item \emph{Quantum Software Implementation and Coding :} Implementation of a quantum software needs writing the required classical and quantum functions to devise the processing logic using high-level quantum programming languages and tools like QCL, Q\#, Scaffold and others. \cite{qcl} \cite{qsharp} The programs written in Domain-Specific Languages (DSLs) are logically synthesized into quantum circuits through logical level schedulers and optimizers. Then, the logical quantum circuit is translated into physical quantum circuit and quantum error correction step is incorporated. These operations on physical qubits generate instructions to be fed to quantum control processor, which further generates hardware specific machine instructions through control pulses.
\item \emph{Quantum System Testing :} The goal of quantum system testing is to ensure that developed system conforms to its requirements laid out in the quantum requirement specification document. Quantum system testing demands quantumness verification and debugging through several quantum verifier models and quantum state reconstruction unlike its classical counterpart. \cite{verification} The statistical model of inferencing in quantum computation (QC) is non-deterministic in nature, which implies a probabilistic measurement of multiple outcomes. Among a set of encoded possibilities, quantum testing will yield the desired outcome with a significantly biased probability, when a quantum module undergoes several iterations. This requires a completely different test-bed preparation and benchmark setting than in classical testing.  \cite{measurement} 
\item \emph{Quantum System Maintenance and Quantum Software Quality Management (QSQM) :} Quantum system maintenance is the final stage of quantum waterfall model representing all modifications and updations made through several performance optimizations like hardware independent algorithmic level optimization and hardware dependent architectural level optimization. This step additionally focuses on quantum software quality management, which can be incorporated in various abstraction levels of development.
\end{itemize}

\section{Quantum Feasibility Study}
Feasibility study, on a classical context is the practical extent to which a project can have a successful completion. Feasibility analysis aims at determining whether the solution considered to accomplish all the requirements is workable in the software. Information such as availability of resources, estimated cost of software development, benefits of the software to the industry after development and cost to be incurred on its maintenance are key factors to be considered during feasibility study. The objective is to establish the reasons for developing the software in terms of user acceptance, adaptability and conformability to established standards. The different types of feasibility considered in the classical SDLCs are technical feasibility and economic feasibility.
In quantum context, putting the first footstep inherently ambiguous due to the lack of definitive answers of questions like short-term and long-term quantum goals. Quantizability of computationally complex business problems and strategically capitalizing the true capabilities of near-term quantum solutions are key concern behind analysis of the feasibility of a quantum project.

\begin{figure*}[t]
  \includegraphics[width=\linewidth,height=16.5cm]{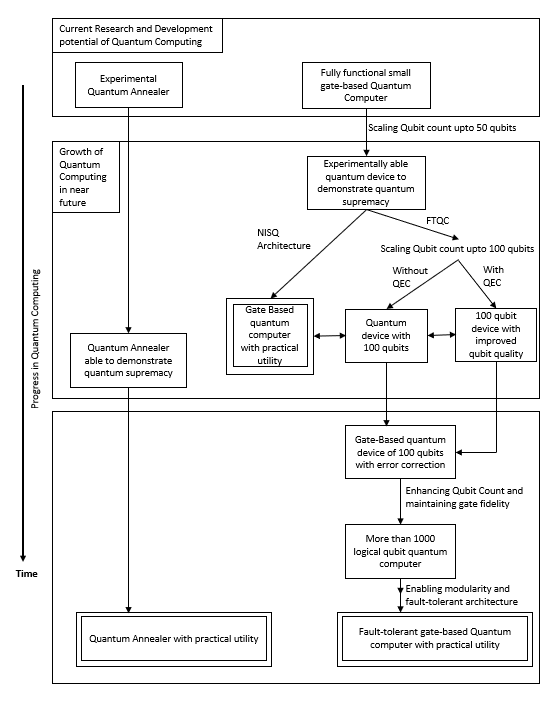}
  \caption {Technical feasibility analysis through stage wise quantum hardware development}
  \label{fig:quera}
\end{figure*}

\begin{itemize}
\item{\textbf{Quantum Technical Feasibility: }
The current state-of-the-art quantum hardware development is in infancy, which necessarily should proceed through several stages such as Component Quantum Computation (CQC), Noisy Intermediate Scale Quantum (NISQ) and Fault-Tolerant Quantum Computing (FTQC). \cite{ftqc} \cite{nisq} CQC demonstrates the basic elements to build up a quantum computer using Physical Machine Description (PMDs) like trapped ions, superconducting qubits, linear photonics, non-linear photonics and quantum dots. \cite{superconducting} \cite{trappedion}\cite{pmd} But, computational capability of CQC is limited by several constraints like scalability, accuracy, decoherence rate to name a few. \cite{scalability} This leads to a second era of quantum computing known as NISQ computing where sufficient availability of physical qubits can demonstrate quantum advantage. Algorithms have to be specifically crafted to fit into NISQ era of computing as these machines support shallow depth of quantum computation such that quantum state preparation and measurement can be accelerated. Mainly, variational algorithm fit in NISQ computer where a quantum-classical hybrid computing is performed and there is a provision to classically vary the experimental parameters. \cite{nisq} In certain computation-intensive applications like realization of Shor’s factorization algorithm, incorporating a huge number of qubits is essential, thus increasing the depth of computation. For these kind of applications, realization of the quantum algorithm in true quantum hardware is very much sensitive to quantum memory, Quantum Error Correction (QEC) and hence technically infeasible in near-term era.}

\item{\textbf{Quantum Operational Feasibility: }
Operational feasibility assesses the extent to which the required software is applicable to solve business problems and meet user requirements. In quantum context, formulating a research and development project with the aim of developing commercial applications for near-term quantum computing will come under quantum operational feasibility study.
\subitem \textbf{Availability of appropriate algorithms: }This includes identifying algorithms with modest problem size and shallow circuit depth to achieve quantum speedup in application domains where classical computers are inefficient. This feasibility study is extended to analyze the hybrid classical-quantum techniques for problems where increase in problem size can dramatically impact solution with huge economic significance. \cite{speedup}
\subitem \textbf{Availability of Skillset: }Achievement of commercial benchmarks through development of a useful quantum application relies on availability of a multidisciplinary pipeline of scientists and researchers. Leveraging true potential of quantum hardware and software needs expertise in multiple areas of science, engineering and technology which in turn requires exploration into quantum physics, mathematics and computing. Lack of quantum computing expertise can affect operational feasibility to a greater extent.
}

\item{\textbf{Quantum Economic Feasibility:}
Economic feasibility determines whether the required software is capable of generating financial gains for an organization, which involves the manpower cost, estimated hardware and software cost, cost of performing feasibility study itself and others. Quantum economic feasibility analysis involves planned resource investment through analyzing quantizability of business problems, quantification and estimation of resources and design of use cases for strategic decision making. \cite{challengeqc} This brings a diverse range of verticals of multi-disciplinary research where a quantum project estimation is performed.}
\end{itemize}

\section{Quantum Requirement Specification}
Feasibility analysis of any quantum project is followed by a quantum requirement analysis phase where the needs and high-level requirements specified in the target project plan (may it be on hardware level or software level) are transformed into unambiguous by definitive, measurable and testable manner, traceable, complete, consistent and stakeholder-approved requirements. Sufficient research and development progress in quantum computing domain is years to achieve and current unavailability of fully-commercial quantum computer yields an abstract overview of quantum requirement analysis and specification. Once technical, operational and economic feasibility have been studied and analysed, quantum requirement specification can be classified broadly into two categories namely, quantum software requirement specification and quantum hardware requirement specification.
\begin{itemize}
\item{\textbf{Quantum Software Requirement Specification: }
A documentation on quantum software requirement specification will consist of all different quantum and classical modules required for successful completion of a quantum project. This mainly will incorporate quantum tools, quantum integrator plugins, logical quantum circuit synthesizer, classical validator and finally classical software requirement specifications.
\subitem \textbf{Quantum Tools:} A functional quantum computer requires extensive software support. This is analogous to the requirements of a classical computation, where different tools are required to support quantum operations including quantum programming languages to enable programmer describing quantum algorithms and quantum compilers to analyze and map the quantum programming language onto quantum hardware setting. These quantum tools offer abstractions that allow programmers to deal with algorithmic level rather than underlined hardware organization and optimization details. Recently available quantum tools are offered by IBM, Microsoft, Rigetti and Google. \cite{superconducting} Microsoft Quantum Developments Kit (QDK) features programming language Q\# as a quantum tool,\cite{qsharp} \cite{trappedion} IBM Quantum Experience is the industry first initiative to build universal quantum computer and thus facilitating quantum software development stack with its software API Qiskit. \cite{qiskit} Rigetti system has offered Rigetti Forest and Cloud Computing Services consisting of quantum instruction language, Quil, and an open source python library for construction of Quil programs called pyQuil, a library of quantum programs called Grove and a simulation environment called Quantum Virtual Machine (QVM). \cite{pyquil} Cirq is another framework provided by Google to create, edit and invoke NISQ circuits. \cite{cirq}
\subitem \textbf{Quantum integrator plugin: }
Integrator plugin is a software component that enhances modifiability of an existing software tool or platform by adding new features to it such that the system design is unaltered due to the modifications. Depending upon the platform with which a quantum module will be integrated, we have divided quantum integrator plugins into two categories – Classical-Quantum Integrator (CQI) and Quantum-Quantum Integrator (Q2I).\newline \textbf{Classical Quantum Integrator (CQI):}
CQI plugin software frameworks are mostly used in optimization and machine learning applications where hybrid quantum-classical computations are to be performed. An integrator plugin system provides a framework which is compatible with any gate-based quantum hardware or simulator. Xanadu proposed a python3 software framework called PennyLane which provide plugins for Strawberry Fields, Rigetti Forest, Qiskit, Cirq and ProjectQ on quantum front and TensorFlow, PyTorch and auto grad on classical front. Each plugin may provide access to one or more devices where devices can be loaded directly from that plugin system without further user intervention. \newline 
\textbf{Quantum-Quantum Integrator (Q2I):}
In quantum computing, software requirement specifications are made based on quantum data generation module and quantum data processing module. Recent research in quantum computer engineering focuses on quantum algorithm development and quantum chip level implementation. Quantum micro-architecture (QuMA) has been proposed which works for a superconducting quantum processor to bridge the gap between quantum software and hardware. This is still in a very nascent phase. \cite{superconducting}
\subitem \textbf{Logical Quantum Circuit Synthesizer: }
Quantum computations are more susceptible to error due to environmental decoherence than conventional classical counterpart. Any arbitrary quantum circuit consists of a cascade of quantum gates, which are further realized using primitive gate operations supported by multiple Physical Machine Descriptions (PMDs). \cite{pmd} Since the cost of executing a quantum operation on a quantum hardware is dependent upon underlying PMD, optimized decomposition of complex quantum operations into simple primitive operations requires a fault-tolerant quantum logical circuit synthesizer. This is another software requirement on algorithmic level, required mostly in applications which are sensitive to error.
\subitem \textbf{Classical Validator: }
In certain applications of quantum computing, an experimental positive outcome of the quantum circuit does not guarantee the correctness of the output. Except the hardest problems known as Bounded-error Quantum Polynomial Complete (BQP-Complete), a claimed quantum solution can prove its correctness by an efficient classical verification procedure. Since, quantum research is in a very nascent stage and requires a lot of practical implementation from the theoretical interest, this software module can be needed in the development of quantum project. \cite{verification}
\subitem \textbf{Classical Software Requirement Specification Module: }
Once the quantum analyst has performed quantum feasibility study followed by the quantum hardware and software requirement specification, a classical analyst will gather all the required information regarding the quantum-classical hybrid software to be developed. By removing al sorts of incompleteness, inconsistencies and anomalies, the requirements should be systematically organized in the form of an Software Requirement Specification (SRS) document, which will contain all the user requirements in structured through informal form.
}
\item{\textbf{Quantum Hardware Requirement Specification: }
Given the current state of quantum technology and recent role of progress, it is highly important to specify the engineering efforts in terms of hardware requirement such that the objective of quantum R\&D project to build a marketable technology can be achieved. Since, performance of quantum computation can be hindered by several technological barriers like insufficient qubit count, low qubit fidelity, qubit error, shorter coherence interval and many more, an unambiguous understanding and declaration of hardware requirements must be specified prior to proceed with the design phase.
\subitem \textbf{Qubit Count: }
Quantum algorithm performs computation by significantly avoiding combinatorial explosion through quantum parallelism. With the help of n-qubit quantum computer, $2^n$ number of computations can be performed in a single step. But this is superficial as we are years ahead to build up a full-scale, fault-tolerant, quantum computer with large number of qubits. In NISQ era of computing, hundreds of physical qubits can be fed for computation which will not be fault-tolerant, but robust to perform computation before decohering. Currently, number of commercially available qubits range around 50 which is sufficiently small from implementing quantum simulation algorithms. \cite{nisq} But from experts’ speculation, if a large quantum computer of 4000 qubits and 100 million quantum gates can be built within next few years, the famous RSA encryption algorithm with 2048-bit keys can be deciphered in seconds. Thus, estimating operational feasibility largely rely on available qubit count.
\subitem \textbf{Quantum Volume: }
Performance of quantum computer does not depend only upon the number of physical qubits operating on a system, but also largely vary on qubit error rate. Number of physical qubits needed to create a single logical qubit for a given Quantum Error Correcting Code (QECC) is largely dependent on the error rate of basic qubit operations. \cite{qecc} Since, various quantum hardware platforms are available today with a wide variety of specifications, there was a need to express the effectiveness of a given quantum hardware by summarizing several capabilities like number of qubits, error rates and qubit connectivity in single metric, known as quantum volume. Specifying quantum volume of the underlying quantum hardware is indeed important as chances of solving complex problems increase with the increase in quantum volume metric. \cite{qubitquality}
\subitem \textbf{Physical Machine Description (PMD): }
Different quantum technologies rely on different physical systems for qubit realization as specific set of primitive quantum operations are required to realize the Hamiltonian of the system. Depending upon different quantum mechanical properties, several PMDs are present which work with different Hamiltonians and its supported operations. \cite{hamiltonian} Generally, PMDs of six quantum systems are mainly taken into consideration, namely Superconducting Qubits (SC), Quantum Dots (QD), Ion Traps (IT), Neural Atoms (NA), Linear Photonics (LP) and Non-Linear Photonics (NP). \cite{pmd} Specifying underlying quantum system and PMD of the quantum hardware are of utter importance as they form the building block of the detailed consideration of specific gate library for quantum synthesis, quantum tools and integrator plugins.
\subitem \textbf{Classical Hardware Processor: }
Classical hardware processor is a host computer responsible for running a conventional operating system with standard supporting libraries to provide all types of software tools and services. This classical processor works synergically with the underlying quantum processor as a support system by creating applications specific to quantum control processors for generating control logic, provide storage and networking services in run-time and many more. Thus, specifying the requirement of a high-performance classical computer with sufficient computation speed will enhance the overall performance of quantum development life cycle by enabling the quantum processor to use its features. These above mentioned quantum requirement specifications will undergo quantum design phase.}
\end{itemize}

\section{Quantum System Design}
In the QDLC model, quantum software design phase comes after quantum feasibility study and quantum requirement analysis have been performed, which further is followed by quantum software coding and implementation phase. Quantum operations are prone to environmental noise and implementing a quantum software incurs a lot of challenges. Thus, a good quantum software design helps to identify the following information.
\begin{itemize}
\item Different number of quantum modules required.
\item Control relationships among both quantum and classical modules.
\item Interfaces among modules in Classical – Quantum (CQ), Quantum – Classical (QC) and Quantum – Quantum (QQ) architecture.
\item Data structures of different modules (Hamiltonian, Hilbert Space, quantum circuits and several Physical Machine Descriptions (PMDs) altogether to represent functionality of several modules) \cite{hamiltonian}
\item Quantum and Classical algorithms required to implement individual modules.
\end{itemize}
Like design phase in classical SDLC, quantum design phase can be divided into two subcategories – quantum software architectural design and quantum software detailed design discussed in the subsections later. Before discussing quantum design activities, it is important to focus on different quantum design methodologies and choosing the optimal one. This requires to first understand and differentiate between analysis and design. Analysis technique aims at elaborating the customer requirements through careful thinking and continuously avoiding making any decisions regarding the exact manner system is supposed to be implemented. On the contrary, the design model obtained from analysis model through several transformations over a series of steps, reflects several decisions taken regarding the exact way system is to be implemented. A good software design is characterized by correctness, efficiency, maintainability and most importantly, understandability of the design to assess system functionality implementability, resource, time and cost optimizations issues, scalability, interoperability and many more.\cite{scalability}

\begin{table*}[ht]
\small
\caption{Different types of coupling exclusively in quantum mechanical system.}
\label{coupling-tab}
\centering
\begin{tabular}{p{0.15\linewidth}| p{0.65\linewidth}}
\hline
Measurement Coupling & \begin{itemize}
\item Exists between a quantum state and environment. Whenever there are any unwanted interactions with the environment, it leads to unintentional measurement of a quantum state.
\item Can be reduced by inserting unitary quantum gates X, Y, Z into a circuit randomly with a specific probability
\item Causes less severity into the performance, as it can be easily modeled by simulating using classical computers.
\end{itemize}\\
\hline
Cross-talk Coupling & \begin{itemize}
\item Caused due to interactions among multiple entangled qubits.
\item Accidental interaction among qubits causes mixing of quantum states and leads to decoherence.
\item Causes unwanted interaction between ancilla qubits and data qubits resulting in degradation of the original quantum state.
\item More severe than measurement coupling, difficult to model.
\end{itemize}\\
\hline
Coherent Coupling & \begin{itemize}
\item Caused due to imprecise classical control of quantum operations.
\item Can adversely impact the quantum algorithms for periodic circuits due to its amplified effect with increase in iteration count.
\item Severe than both measurement and cross-talk coupling as classically simulating a quantum system of size n requires an exponential computation time.
\end{itemize}\\
\hline
Non-coherent Coupling & \begin{itemize}
\item Caused by loss of energy from a quantum state into the environment spontaneous emission of photon.
\item Exists in quantum system with multiple energy levels.
\end{itemize}\\
\hline
\end{tabular}
\end{table*}

\subsection{Quantum System Architectural Design}
This design subphase focuses on functionalities and interactions of and between quantum software and hardware components like quantum tools, quantum integrator plugins with respect to both classical and quantum modules, logical quantum circuit synthesizer, underlying PMDs, classical hardware etc. \cite{qcomplexity} The intrinsic complexity of quantum system makes it extremely hard to set the benchmarks for a good quantum architectural design. Despite the unavailability of available research resources and compacted system patterns, we have identified several good quantum architectural design characteristics to measure and decide both the functional strength of a quantum module (cohesion) and measure of the degree of interdependence between two quantum processing units (alternatively, analogous to coupling) Along with different types of cohesion or degrees of freedom existing in classical SDLC like coincidental, functional, logical, temporal, procedural, communicational and sequential cohesion, a quantum system primarily possesses two different degrees of freedom, namely – quantum temporal cohesion and quantum spatial cohesion.

\begin{table*}[ht]
\small
\caption{Generic computing pattern for a quantum algorithm on gate model of computing.}
\label{algo-gate}
\centering
\begin{tabular}{p{0.15\linewidth}| p{0.65\linewidth}}
\hline
Initialization of Registers &
Setting up and initializing the quantum registers and classical registers required for computation.\\ \hline
Uniform Superposition &
Applying Hadamard transform to achieve an uniform superposition  of quantum states.\\
\hline
Entanglement &
Entangling multiple qubits to form an entangled quantum state.\\ \hline
Creation of function table &
Analyzing quantum algorithm in a query model in order to compute boolean functions, where inputs are given in a black box aiming to compute the function value for an arbitrary input string with fewer number of queries.\\ \hline
Oracle Setup &
Setting up an oracle, which us a black box operation and can be used in computation which is obtained as input from another quantum algorithm.\\ \hline
Uncompute Task &
Applying reversible computing logic to remove entanglement from resulting state.\\ \hline
Phase Shift &
In order to identify and mark a single quantum state from the superposed set of quantum states, controlled phase shift operation is performed using unitary quantum gates.\\ \hline
Amplitude Amplification &
A tool of choice for quantum algorithm designers such that the success probability of query algorithms can be increased.\\ \hline
Speedup via verification &
Performing quantum verification and achieving substantial speedup. \\ \hline
Classical Measurement &
Splitting up the quantum and classical states by performing a classical measurement.
\\
\hline
\end{tabular}
\end{table*}

\subsubsection{Quantum Temporal Cohesion}
Quantum computing obeys laws of quantum physics and achieves exponential computing power due to quantum superposition. While a state in quantum is in superposition of several mathematical possibilities, it will allow the operation to be executed on multiple qubits in the same time span. When a single qubit operation performs a task on a cluster of qubits under superposition, that quantum operation is said to exhibit temporal cohesion. Achieving temporal cohesion depends on qubit count and suitable problem encoding but once achieved can affect the degree of non-determinism of quantum software over classical counterpart in a significant extent.
\subsubsection{Quantum Spatial Cohesion}
A quantum module ( or circuit ) is to possess spatial cohesion, if there exists a correlation between neighboring qubits in order to interact. Simulation of a quantum algorithm can be performed in a pseudo quantum environment, but its implementation involves realization of quantum circuits through physical synthesis of quantum gates. This type of cohesion in a quantum system can be observed by looking into the constraints imposed by the methodological framework for physical synthesis regarding placement of qubits and quantum operators. If two interacting qubits are non-neighboring, they will be more susceptible to environmental decoherence resulting in consequent failure in computation. This is spatial cohesion, due to which, we have to apply SWAP gates and transform Non-Linear Nearest Neighbors (Non-LNN) architecture to Linear Nearest Neighbor (LNN) architecture. \cite{ghosh}
\subsubsection{Quantum Coupling}
In classical context, the term coupling is used to represent the degree of interdependence between different modules. The degree of coupling between two modules is determined by their interface complexity. A highly coupled system infers a weak design aspect, as complexity of parameters of one module can affect the performance of the other. Different types of coupling, in increasing order of their severities include data coupling (communicating using elementary data item), stamp coupling (communicating using a composite data item or a data structure), control coupling (using data from one module to direct the execution of the other), common coupling (sharing global data items) and content coupling(sharing an entire code). There are the significant benchmarks on coupling to determine the efficiency of a software design in classical computing. In quantum, the performance of a single quantum mechanical system often undergoes several complications due to different types of coupling or interdependence among qubits. We have coined four several dependencies existing in today’s quantum devices, namely – measurement coupling, cross-talk coupling, coherent coupling and non-coherent coupling as shown in table \ref{coupling-tab}.

\begin{table*}[ht]
\small
\caption{Generic computing pattern for a quantum algorithm on adiabatic model of computing.}
\label{algo-adia}
\centering
\begin{tabular}{p{0.15\linewidth}| p{0.65\linewidth}}
\hline
Beginning Hamiltonian ($H_B$) &
Stating the initial Hamiltonian whose ground state is easy to create. \\ \hline
Problem Hamiltonian ($H_P$) &
Defining a problem Hamiltonian whose ground state will encode the solution to the optimization problem. \\ \hline
Choosing Hamiltonian Spectrum ($s$) &
Defining gap between successive energy states.\\ \hline
Run time estimation &
Time evolution needed to achieve required adiabaticity. \\ \hline
Slow interpolation &
Interpolating from $H_B$ to $H_P$ with a smoothly varying Hamiltonian $H(s), s \in [0,1]$ .\\ \hline
Error analysis &
Unitary control error, errors in interpolation and final Hamiltonian, high frequency error and thermal noise are taken into consideration. \\ \hline
Efficiency checking &
Inverse polynomial speedup refers to efficient quantum algorithms. \\ \hline
\end{tabular}
\end{table*}

\subsection{Quantum System Detailed Design}
Detailed design in QDLC will focus on the problem statement, processing logic, quantum algorithms based on that processing logic, data structures and semantics for data definitions. Quantum algorithm are able to solve certain computation tasks with exponentially fewer number of steps than the existing best known classical algorithm. For example, finding number of prime divisors an $n$-bit number $N$ requires $exp(O(n^{\frac{1}{3}}))$ steps by the best classical algorithm, which on the contrary can be solved in $O(n^2 \log n)$ time in quantum context using Shor's quantum algorithm for prime factorization. \cite{shor}

Despite this super polynomial power of quantum computing, it is important to realize that quantum computers can not solve all the problems involving combinatorial explosion if the algorithm lacks the detailed problem structure. Moreover, the current era of quantum computing is based on noisy intermediate scale quantum (NISQ) devices. \cite{nisq} Hence, developing quantum algorithms suited for NISQ devices is of utter importance for real-time implementation. Considering the power of quantum algorithms on an ideal quantum hardware, it is important to analyze the generic computation steps or computing patterns to run a quantum algorithm on both gate model of computing and adiabatic model of computing. In the following table \ref{algo-gate} and table \ref{algo-adia} the major building blocks for a generic quantum algorithm on gate model and adiabatic model are shown respectively.

\section{Quantum Software Coding and Implementation}
In classical SDLC, coding stage is performed once design stage is complete and the design documents are reviewed successfully with an aim to transform the design of a system into user understandable code in a high–level language. Unlike classical, programming and implementing a quantum system requires a different style of operations and concepts, requiring new programming languages, frameworks and tools. Designing a quantum algorithm requires a mathematical construct to solve a given problem, which has to undergo quantum programming followed by a series of complier transformations and optimizations before executing on a real hardware. The process of implementation begins with a high–level programming abstractions, where quantum algorithms are programmed using high–level domain specific languages (DSLs). High–level QC programming languages can be categorized into several genres like functional, imperative, embedded and strongly – typed quantum languages as shown in table \ref{qprogram-tab}.

\begin{table*}[ht]
\footnotesize
\caption{Evolution of quantum programming languages (from high-level abstractions to hardware specific implementation)}
\label{qprogram-tab}
\centering
\begin{tabular}{p{0.10\linewidth}| p{0.15\linewidth}| p{0.55\linewidth}}
\hline
\multirow{4}{*}{}
High-level QC programming languages (DSLs) & Functional Languages (Q\#, Quipper, QuaFL, LIQUi|>) & 
Compact language constructs.
Less error prone than imperative languages.
\newline Suited for FTQC era.
\\ \hhline{~--}
& Imperative Languages (QCL, Scaffold, ProjectQ) & 
NISQ-compatible language constructs.
\newline Resource-efficient and less compact than functional QC programming tool kits.
\\ \hhline{~--}
& Embedded Languages & 
Embedding current QC language in existing non-QC language.
Extension of base language.
\newline Allowance to use base language's software stack.
\newline Accelerated implementation.
\\ \hhline{~--}
& Strongly typed QC language & 
Stricter code inspection by setting strong type checking rules.
\newline Assuring type safety and software reliability.
\\ \hline
\multirow{2}{*}{\newline  \newline }
Intermediate-level assembly languages (QIR) & Quantum Assembly Language (QASM) & 
Limited to basic quantum operational constructs.
\newline Expressing a simple quantum circuit as a gate sequence.
\newline Lack of Scalability.
\newline Limited expressive power.
\newline No construct for iterative procedures and sub-routine calls.
\\ \hhline{~--}
& Open QASM &
Combining the features of classical assembly languages and C with existing constructs for QASM
\\ \hline
Low level languages and frameworks & Python framework QcoDeS (data acquisition and interaction with physical qubits), OpenQASM backend for IBMQ, ARTIQ (Ion Trap) & 
Translating openQASM into specific control instructions. 
\newline Generating code for control processor.
\newline Driving signals from control processor to control and measurement plane.
\\ \hline
\end{tabular}
\end{table*}

\begin{figure*}[t]
  \includegraphics[width=\linewidth,height=14cm]{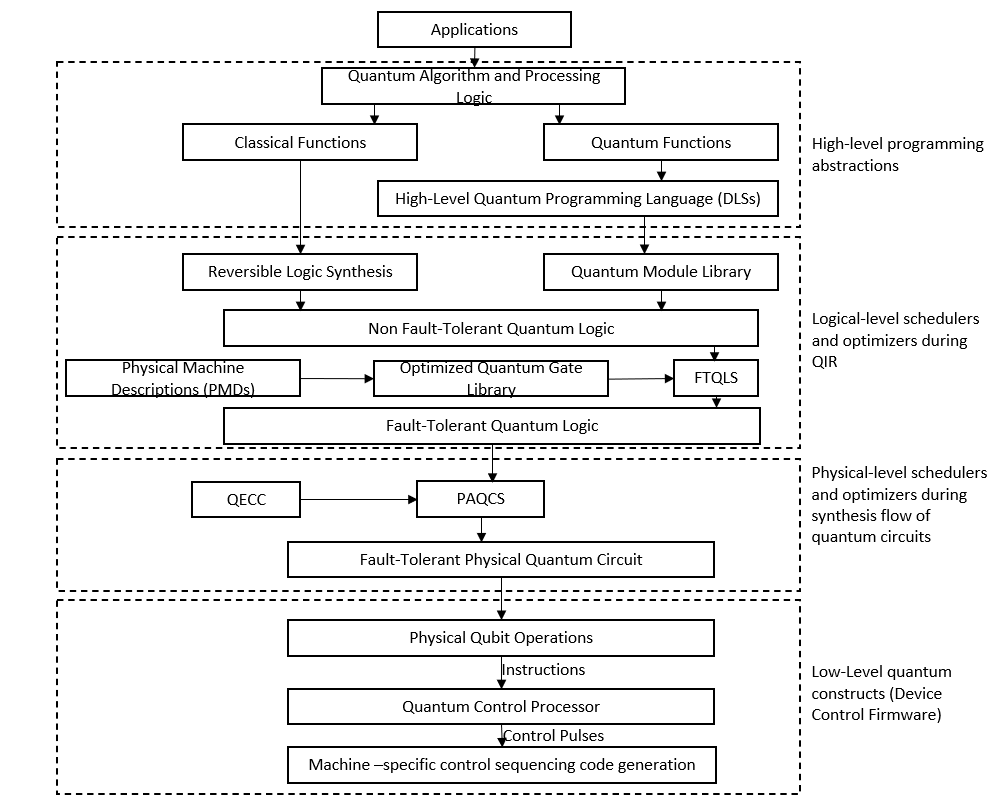}
  \caption {Quantum module implementation (series of compiler transformations and optimizations to translate high-level quantum program into specific hardware instructions)}
  \label{fig:qimplement}
\end{figure*}

Functional programming languages like Q\#, Quipper, QuaFL, LIQUi|> are more compact and less error prone, hence are considered to be more suited for FTQC era. \cite{ftqc} Q\# is a Quantum Development Kit (QDK) developed by Microsoft, which is enabled with the feature of creating qubits, entangling them and performing superposition of qubits via unitary quantum gates like controlled NOT, Hadamard, Pauli X, Y, Z and many more. Q\# is expected to implement quantum operations in the form of topological qubits. The currently available quantum simulator with QDK as Q\# is capable to process upto 32 qubits on user machine, expandable upto 42 qubits on Azure platform.\cite{qsharp} Quipper is a scalable, embedded functional programming language that provides a high – level circuit descriptions including gate–by–gate descriptions of circuit fragments and powerful operators to assemble and manipulate those circuits. Syntactic style of Quipper follows a mix of procedural and declarative programming with a built–in facilitate to automate reversible quantum circuit synthesis and provides support for different phases of execution like during compile time, circuit generation time and circuit execution time. It allows dynamic parametric circuit generation and hence provides an additional library support for several quantum functions like quantum integer and fixed–point arithmetic, Quantum Fourier Transform (QFT), Quantum Random Access Memory (QRAM) implementation, pseudo–classical circuit simulations, exact and approximate decomposition of quantum circuits into a sequence of quantum gates to name a few. QuaFL is a statically typed DSL used for writing high level definitions of algorithms in order to be compiled into logical quantum circuits. QuaFL allows programmers to make a use of high-level data structures to distinguish between classical and quantum sections of a program and ensure physical realizability of the quantum section with the help of orthogonality checking algorithm. Its ability to provide type size annotations, helps to facilitate automated computation of necessary quantum resources. LIQUi|> is another functional quantum software architecture and tool suite for quantum computing developed by Microsoft research, which includes programming language, optimization and scheduling algorithm. Its extended functionality allows Hamiltonian simulation, quantum noise model simulation and supports cloud operation. Specific quantum algorithms that can be simulated with LIQUi|> are simple quantum teleportation, Shor’s integer factoring algorithm, molecular ground state energy computation, quantum error correction, quantum associative memory and quantum linear algebra. \cite{hamiltonian}

Imperative quantum programming languages like QCL, Scaffold, ProjectQ are mostly suited for NISQ era of computing. Quantum Computation Language (QCL) is one of the first implemented programming language in quantum. \cite{nisq} Its syntax resembles the syntax of classical C programming language and different type of standard quantum operators supported by QCL standard library include Controlled-NOT (CNOT) with many target qubits, multi-qubit Hadamard operation, phase and controlled phase operation, basic quantum algorithms for modulo addition, multiplication, exponentiation and QFT. \cite{qalgo} Scaffold is another high-level, imperative quantum programming language developed to provide extended services to C by introducing new data types, qbit and cbit, obtained as a result of measuring a quantum bit and a classical bit respectively. Scaffold first introduced a class of modules known as Classical-To-Quantum-Gate (CTQG) to allow sub-circuits to be defined as classical logical circuits. \cite{qalgo} Another imperative open-source quantum software framework is ProjectQ, which allows users to implement a quantum program in python with complex syntax and then translates the programs to any type of backend. Further advancement of high-level quantum DSLs leads to embedded and strongly typed quantum programming languages with additional features like quick design due to no “from-scratch” concept, strong type checking rules, greater reliability and many more. Unlike gate model of computing, D-Wave 2000Q system is built up based on adiabatic model of computing and provides a standard internet Application Programming Interface (API) for end-users. Users are allowed to access the system either in cloud platform or high performance computing (HPC) integrated platform with the help of available client libraries in C, C++, Python and MATLAB. D-Wave's open source Ocean software development kit (SDK) allows developers to implement their own algorithms and applications within their existing environment. \cite{qalgo} A user program submitted to D-Wave system will represent a set of values which are mapped to the weights of qubits and the strength of couplers. \cite{adiabatic} The system processes the set of values and other user-specified parameters together to send a single quantum machine instruction (QMI) to the quantum processing unit (QPU). QPU is initialized into ground state of a known Hamiltonian which undergoes a process of annealing to remain in the ground state during the entire adiabatic time evolution. Each qubit will yield a 0 or 1 at the end of computation and the final state will represent the optimal or near-optimal solution to the given problem. Qbsolv is a tool developed by D-Wave that takes large Quadratic Unconstrained Binary Optimization (QUBO) problem and partitions it into smaller QUBOs. This technique resolves the scalability issue as smaller sized QUBOs can easily suffice to fit into the system capacity and topological constraints of D-Wave quantum processor. \cite{hamiltonian} All these user operations are performed via “command-line interface” through web over cloud service, due to commercial and physical unavailability of quantum systems, which can further evolve with the advancement in overall quantum development ecosystem. \cite{qalgo}

Quantum system implementation undergoes a synthesis step, where intermediate-level quantum assembly languages alternatively referred as Quantum Intermediate Representation (QIR) are used to generate instructions for quantum control processor. The quantum algorithm programmed in any of the DSLs is first translated into a quantum circuit which is composed of a cascade of quantum gates. This logical level synthesis is referred as Non-FT (Non fault-tolerant) quantum logic. The quantum gate library obtained from non-FT logic undergoes Fault-Tolerant Quantum Logic Synthesis (FTQLS) where quantum gate library is optimized by decomposing the multi-qubit quantum gates into simple one-qubit and two-qubit gates based on underlying PMDs to generate a fault-tolerant quantum logic. \cite{ftqls} A fault-tolerant physical quantum circuit is synthesized by applying quantum error correction \cite{qecc} (QEC) and physical design Aware Quantum Circuit Synthesis (PAQCS) to transform a non-LNN architecture into LNN architecture. In the last phase of compilation, intermediate-level languages (QIR) represented using Open QASM are translated into control instructions which is fed to quantum control processor to generate the control pulses. \cite{qasm} \cite{paqcs} \cite{ghosh} These control pulses are the final low-level quantum constructs wich generate machine-specific control sequencing code. The detailed phase of compilation and optimization involved in implementation of a quantum software while translating a high-level quantum program written in DSLS into specific hardware instructions executable on a real quantum device are shown in figure \ref{fig:qimplement}.

\section{Quantum Testing}
Specification, verification and debugging of quantum modules or programs are inherently complex processes due to underlying difficulty in order to make an error-free design by quantum computing software and hardware. The intractability of quantum simulation also puts a limit in the amount of available predesign testing and simulation suite.  \cite{verification} Moreover, conventional debugging tools based on measuring program variables during execution would disrupt the process and hence are not suitable as measurement causes collapse of a quantum state. \cite{measurement}

Unlike classical computing and classical software verification, quantum verification aims at answering the question, is it possible to verify for a classical client the answer provided by a quantum computer?
\subsection{Challenges of Quantum Verification}
The difficulties in answering the questions asked in quantum verification or quantumness testing process stem from fundamental principles of quantum mechanics and appear inherently unattainable.
\begin{itemize}
\item Computing power deficiency of classical verifier
Due to exponential power of quantum systems, direct simulation of a moderately scaled quantum devices by a classical computer is impossible to exercise.
\item Inherent randomness of quantum mechanical system
Quantum mechanical laws severely limit the amount of information about a quantum state that could have been obtained due to collapse action caused by measurement of the quantum state.
\end{itemize}
\subsection{Quantum Verifier and Debugging Tools}
Exploring the theory of interactive proof systems and its deep interaction with classical cryptography have led to various ideation of quantum debugging process. This yields introduction of several models based on the nature of verifier and its interactions with the quantum system.
\begin{itemize}
\item ‘Slightly Quantum’ verifier model:
This model incorporates a ‘slightly quantum’ experimentalist or verifier who has the ability to manipulate a constant number of qubits and has the access to quantum channel to the quantum computer. This model of quantumness verification is used in quantum authentication techniques to help maintaining the quantum computer honest.  \cite{verification} \cite{verifier} Security proofs of such protocols are extremely delicate.
\item Classical verifier and adversarial multiple quantum device setting model:
This model considers a classical verifier interacting with multiple quantum devices through shared entanglement. This describes a scheme for efficient characterization of quantum devices and verification of their answers. This model, alternatively known for quantum cryptographic context where device independence is studied in adversarial quantum setting.  \cite{verification} This model is used in certification process of quantum random number generation and used as a tool for generation of protocols in order to obtain fully device independent Quantum Key Distribution (QKD).
\item Classical verifier and single quantum device model:
This model considers a classical verifier interacting with a single quantum device, where post-quantum cryptography has been used to keep the device honest. Research work on this model has shown the process of efficient verification of quantum supremacy based on trapdoor claw-free function implemented under learning with errors (LWE). Another work has shown the use of trapdoor claw-free function by a classical client to delegate a computation to a quantum computer in the cloud without compromising the privacy of its data, which is a base for quantum fully homographic encryption. An ingenious protocol based on trapdoor claw-free function can effectively verify the output of a quantum computer. This is done through allowing a classical computer to interactively verify the result of an efficient quantum computation by constructing a measurement protocol, which enables a classical verifier to use a quantum prover as a trusted measurement device. \cite{measurement}
\end{itemize}
\subsection{Quantum testing through quantum state reconstruction}
Measurement process changes the state of the quantum mechanical system, hence understanding the error seeds through measuring the state of a quantum computer is indeed a difficult task. Each measurement yields only a single index of the quantum state. This causes quantum state reconstruction through repeated preparation (P) and measurement (M) to generate the full-scale probability distribution of the quantum state under measurement. \cite{measurement} This repeated reconstruction of quantum states is known as quantum-state tomography, which provides an estimate for the underlying quantum state. Tomography process represents a complete description of the errors caused during a circuit operation but its implementation requires and extremely large number of steps to characterize the circuit transformation from the input quantum state to output quantum state for an n-qubit system, $2^{2n}$ number of measurements are used for ensuring adequacy in the number of samples in each possible output state. 
\section{Quantum Software Quality Management}
\begin{figure*}[t]
  \includegraphics[width=\linewidth]{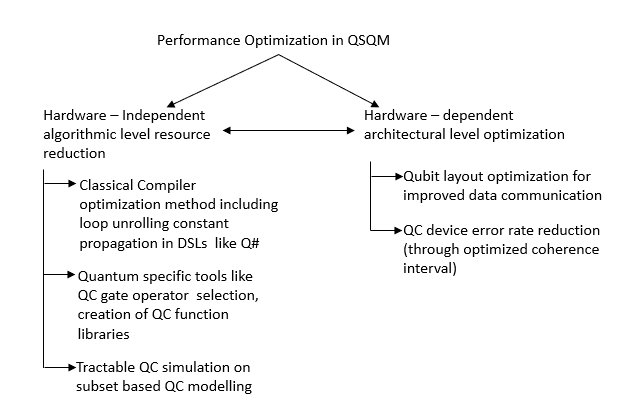}
  \caption {Performance optimization in Quantum Software Quality Management (QSQM)}
  \label{fig:qsqm}
\end{figure*}

Compilation process via compiler tools helps to map a quantum algorithm down to a quantum hardware. Often, hardware availability of quantum system is preceded by development of a compiler system coupled with features like resource estimation and simulation. Quantum Software Quality Management (QSQM) primarily can be thought of as generating focussed optimization in order to reduce the estimated runtime of a quantum algorithm from a high-degree polynomial to a low-degree polynomial as shown in Figure \ref{fig:qsqm}. This optimization can have a dramatic impact on the resources required for execution of quantum computation and can bring down the expected time to solution from billions of years to hours or days. Compilers, in both classical and quantum regime perform multiple resource optimizations while analyzing and translating an algorithm to machine executable code. Successful quality management through optimization can indeed accelerate the arrival process of quantum and put the classical system closer to tipping point by significantly reducing the required number of qubits and amount of time required to execute an algorithm. 

Moreover, the digital Noisy Intermediate Scale Quantum (NISQ) systems are prone to sensitivity to the quality and efficacy of a quantum software ecosystem. Making an effective use of NISQ machine necessities a fine-tuned optimization process due to the resource constrained nature of NISQ Systems with limitation in number of qubits and low gate fidelities. Full stack information flow is required to identify tractable mappings from algorithm design phase to implementation. \cite{nisq} Algorithm and mapping choices are often influenced by noise or error characteristics and level of achievable parallelism, causing the need of layer to layer communication and complex system design. This drives specific aspects of toolchain design like limiting cross-layer abstraction, encouraging the use of libraries of hand-tuned modules to name a few. A deep insight into QSQM will drive research in algorithms, device technologies and analysis of optimized designs with a remarkable cost-benefit tradeoff. Quantum SQM activities primarily encompasses improvisation in NISQ circuit width (qubit count) and circuit depth (number of time steps or operation count).
\subsection{Hardware – Independent algorithmic level resource reduction}
Large scale quantum computation is still in its nascency due to challenges of executing a quantum algorithm on real quantum hardware or simulating QC systems at scale. But, in contrast to QC simulation or real machine execution, analysis of resources can be made a priori to execution as it requires only the time and resource consumption that would be required to answer the feasibility. Algorithm resource analysis does not compute the answer itself, so it does not need to compute the full quantum state information and hence is able to analyze algorithmic performance for upto hundreds of thousands of qubits and millions of quantum operations or execution time steps. \cite{qhardware}
\subsubsection{Classical optimization for Domain – Specific Language (DSLs)}
A dominant approach to programming a quantum computer requires an existing high-level language (Commonly known as DSLs) with libraries to allow for the expression of quantum programs. In order to address the need for expressivity and abstraction, DSLs are designed with classical compilation support to provide a clean separation between the code executed in the quantum context from the driver code which is further programmable in traditional high-level language. Classical compiler optimization tools used for reducing resources include function inlining, dead code elimination, common subexpression elimination, loop invariant code motion, loop unrolling, constant propagation and many more. Figure \ref{fig:qsqm} represents an analytical study on the impact of compiler optimization in high-level languages like python which is a base for embedding of quantum DSLs.

\subsubsection{QC gate operator selection}
The role of physical level compilation stage involves selection and synthesizing particular gate functions needed for computation, which are comparable to the instruction set architecture (ISA) of a conventional computer. \cite{qhardware} Gate Synthesis includes decomposition of single qubit operations into sequences of elementary quantum gates to enable a general quantum circuit expressed in arbitrary unitaries to be synthesized into an approximate quantum circuit composed of a sequence of elementary gates. \cite{adiabatic} The cost of most frequently used gates in quantum computation can be analyzed deeply by counting the sequence of basic physical operations and presenting the upper bound of the cost function in the worst case. Number of required quantum gates increases gradually with the increase in accuracy demand. State-of-the-art synthesis methods have been developed to enable a quantum single-qubit rotation to be synthesized with the help of roughly $log(\frac{1}{\epsilon})$ number of gates where $\epsilon$ represents accuracy of the gate sequence. \cite{coherencequbit}
\subsubsection{Traceable QC Simulation on subset-based QC modelling}
The fundamental challenges of performing QC simulation to improve the quality of quantum software development process are speed and scalability of quantum space. \cite{scalability} But, if QC simulators can be built to model subsets of quantum operations, process becomes tractable and error correction can be analyzed on more than thousands of qubits. Example can be simulation of Toffoli circuits which contain Pauli-X, Controlled NOT and Controlled-Controlled NOT operations, which can be efficiently simulated on classical inputs and enables efficacy in studying and debugging large-scale arithmetic quantum circuits. This quality improvement technique is generally used by the developer to provide a classical implementation of the quantum oracle function and hence achieving a higher level of abstraction. \cite{coherencequbit} Another example of subset-based QC modelling can be thought of as computation of a mathematical function like modular addition on a quantum computer, where simulator can simply implement modular addition on each of the computational basis states rather than to apply a set of sequenced quantum operation required for reversible modular addition.
\subsection{Hardware – Dependent Qubit-layout optimization and QC device error rate reduction}
In the design phase of QDLC, an abstract layered architecture model of QC hardware has been illustrated whose layers are quantum data plane, control and measurement plane, control processor plane and classical host processor. Components of a quantum system are fragile in nature and its interaction with the environment causes the information stored in the system to decohere resulting in error and consequent failure of computation. \cite{challengeqc} Moreover, the speed of a quantum computer can never be faster than the time required to create the precise signals needed to perform the quantum operations. As a matter of quantification of the fact regarding speed of a quantum gate operation, it is currently tens to thousands of nanoseconds for superconducting qubits using microwave and low-frequency electric signals and one to a hundred microseconds for trapped ion qubits using forms of electromagnetic radiation like optical signals.\cite{superconducting} \cite{qhardware} \cite{basicqc}
\subsubsection{Optimizing quantum data plane}
\begin{itemize}
\item Enhancing qubit fidelity
\subitem Strong isolation from outer environment.
\subitem Use of cryogenic CMOS, single-flux Quantum (SFQ), Reciprocal Quantum Logic (RQL) and adiabatic quantum flux parametrons.
\item Controlling the effect of limiting qubit connectivity 
\subitem Consideration of architectural constraints during mapping the inter-qubit connectivity.
\item Increasing qubit coherence time
\subitem Removal of ambient noise from magnetic field fluctuations and reduction of phase noise from local oscillators.
\subitem Use of improved materials.
\subitem Reduction in sensitivity of quantum-devices by using symmetrical design and operation.
\subitem Adoption of dynamic methods like ‘spin echo’ and ‘geometric manipulations’ to counteract quantum decoherence.
\item Increasing qubit capacity in a single quantum data plane module
\subitem Adoption of highly specific mode of transmission to affect the desired qubit without state alteration of other qubits.
\subitem Clever placement of physical qubits in n-Dimensional architecture.
\end{itemize}

\subsubsection{Optimizing control and Measurement Plane}
\begin{itemize}
\item Minimizing qubit manufacturing error through preventive Quantum Error Mitigation (QEM)
\subitem Application of composite pulses to reduce systematic errors.
\subitem Use of dynamical decoupling sequences to reduce coherent dephasing errors.
\item Minimizing signal crosstalk errors.
\subitem Reduction of qubit interdependence by using control pulse shapes.
\subitem Use of periodic system calibration for error mitigation.
\end{itemize}
\subsubsection{Optimizing control processor plane}
\begin{itemize}
\item Identifying and triggering proper Hamiltonian or sequence of quantum gate operations.
\subitem Development of scalable shallow circuit (in terms of circuit depth) for algorithm implementation.
\item Reducing overhead in applying Quantum Error Correction (QEC) algorithm.
\subitem Time synchronization between error correction operations and quantum oracle operations while measurement.
\item Scaling control processor planes to support large quantum machines.
\subitem Integration of a stand-alone QC with High Performance Computing (HPC) system.
\subitem Progression toward sophisticated accelerator model by introducing tightly coupled interconnected system of multiple CPUs with one or more Quantum Processing Units (QPUs).
\end{itemize}
\subsubsection{Optimizing host processor}
\begin{itemize}
\item Optimization in terms of storage and networking services to back-up quantum operations.
\end{itemize}

In quantum context, quality management can be incorporated in various abstraction levels, starting from compilation of the algorithm to the full-scale hardware implementation. \cite{qhardware} \cite{basicqc} This might lead to an iterative optimization process where sometimes circuit width and depth are analyzed after the algorithm has been mapped to a discrete set of sequences of single qubit and two qubit operations to understand the trade-off between resource requirement and operational accuracy or post QEC performance analysis to have an estimated account of QEC and communication overhead.

\section{Near Term Quantum Solution and Challenges at Hand}
One of the major differences between a classical computer and a quantum computer is in
how it handles small unwanted variations, or noise, in the system. Since a classical bit is either one or zero, even if the value is slightly off (some noise in the system) it is easy for the operations on that signal to remove that noise. \cite{challengeqc} In fact, today’s classical gates, which operate on bits and are used to create computers, have very large noise margins as they can reject large variations in their inputs and still produce clean, noise-free outputs. Because a qubit can be in any combination of '0' and '1', qubits and quantum gates cannot readily reject small errors (noise) that occur in physical circuits. As a result, small errors in creating the desired quantum operations, or any stray signals that couple into the physical system, can eventually lead to wrong outputs appearing in the computation. \cite{challengeqc}

Although the physical qubit operations are sensitive to noise, it is possible to run a
quantum error correction (QEC) algorithm on a physical quantum computer to emulate a
noise-free, or “fully error corrected,” quantum computer. Without QEC, it is unlikely that a complex quantum program, such as one that implements Shor’s algorithm, would ever
run correctly on a quantum computer. However, QEC incurs significant overheads in
terms of both the number of physical qubits required to emulate a more robust and stable
qubit, called a “logical qubit,” and the number of primitive qubit operations that must be performed on physical qubits to emulate a quantum operation on this logical qubit. While QECs will be essential to create error-free quantum computers in the future, they are too resource intensive to be used in the short term: quantum computers in the near term are likely to have errors. This class of machines is referred to as noisy intermediate-scale quantum (NISQ) computers. \cite{nisq}

While a quantum computer can use a small number of qubits to represent an exponentially
larger amount of data, there is not currently a method to rapidly convert a large amount of classical data into quantum states (this does not apply if the data can be generated algorithmically).
For problems that require large inputs, the amount of time needed to create the input quantum state would typically dominate the computation time, and greatly reduce the quantum advantage.

Measuring the state of a quantum computer “collapses” the large quantum state to a single
classical result. This means that one can extract only the same amount of data from a quantum computer that one could from a classical computer of the same size. To reap the benefit of a quantum computer, quantum algorithms must leverage uniquely quantum features such as interference and entanglement to arrive at the final classical result. Thus, achieving quantum speedup requires totally new kinds of algorithm design principles
and very clever algorithm design. \cite{speedup} Quantum algorithm development is a critical aspect of the field.

\section{Conclusion and Future Scope}
Typically, any technological progress can be predicted by extrapolating upcoming trends from available dataset with the help of quantifiable metric of measuring progress. Existing commercial utilities in quantum machines and algorithms have already drawn attention of thousands of investors to invest in this billion dollar quantum industry. This has made quantum computing not to be just a theoretical concept but a palpable for enterprises and has a great potential in delivering business value. In this paper, we have analyzed several operational constraints, implementation challenges, research trends in both existing NISQ era and upcoming FTQC era. Our proposed infrastructure for quantum software development through a generic QDLC waterfall model is able to structurally strengthen and facilitate the initial research growth in near term quantum era. This delineate paper can indeed provide a direction for shaping the near-term business trajectory into quantum through substantial growth in research and development.

\begin{IEEEbiography}[{\includegraphics[width=1in,height=1.25in,clip,keepaspectratio]{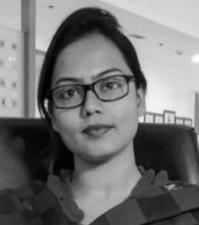}}]{Nivedita Dey} has full time association with the first quantum research and development lab of Kolkata, India. Prior to this, Nivedita had worked for an Israel based company with a designation of Quantum Software Researcher and also made her contribution in academia as Assistant Professor in Computer Science. Her research interests include quantum algorithm design, quantum circuit synthesis and quantum information processing. She is presently involved as one of the key persons in development of a quantum based cryptographic suite and in designing its use cases in varied domain of industry and academia. 
\end{IEEEbiography}

\begin{IEEEbiography}[{\includegraphics[width=1in,height=1.25in,clip,keepaspectratio]{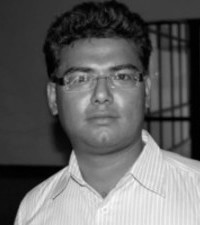}}]{Mrityunjay Ghosh} is currently associated with University of Calcutta and leading research and development of industry first quantum lab of eastern India. Previously, he was head of quantum technologies of an Israel based company and worked  as Assistant Professor of different state level Universities. He has 11 years of experience in industry, research and academia. His research areas of interest include reversible computing, quantum algorithm, quantum circuit simulation and synthesis. Presently, he is leading a project on development of a quantum based cryptographic suite using quantum random number generator.
\end{IEEEbiography}

\begin{IEEEbiography}[{\includegraphics[width=1in,height=1.25in,clip,keepaspectratio]{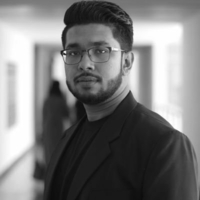}}]{Subhra Samir Kundu} is a post graduate student in Computer Application and currently associated as project intern in the research and development wing of QRDLab. His research interests include quantum circuit simulation and quantum machine learning. He is focused to contribute in quantum computation area for development of new quantum engineering principles and algorithms.
\end{IEEEbiography}

\begin{IEEEbiography}[{\includegraphics[width=1in,height=1.25in,clip,keepaspectratio]{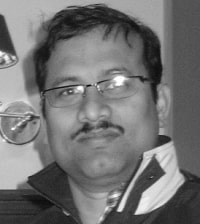}}]{Amlan Chakrabarti} is a Full Professor in the A.K.Choudhury School of Information Technology at the University of Calcutta. He was also the Dean, Faculty of Engineering and Technology of his university. He was a Post-Doctoral fellow at the School of Engineering, Princeton University, USA during 2011-2012. He has almost 20 years of experience in Engineering Education and Research. He is the recipient of prestigious DST BOYSCAST fellowship award in Engineering Science (2011), Indian National Science Academy (INSA) Visiting Faculty Fellowship (2014), JSPS Invitation Research Award (2016) from Japan, Erasmus Mundus Leaders Award from European Union (2017), Hamied Visiting Professorship from University of Cambridge, UK (2018) and Siksha Ratna Award by Dept. of Higher Education Govt. of West Bengal (2018). He has also served in various capacities in various higher education organizations both at national and international levels.
\end{IEEEbiography}

\end{document}